\documentclass{article}
\usepackage[utf8x]{inputenc}
\usepackage[hmarginratio=1:1,top=1.2cm,columnsep=20pt]{geometry}
\usepackage{booktabs,float,hyperref,paralist,multicol,titlesec}
\newcommand{\HRule}{\rule{.5\linewidth}{0.5mm}}
\newcommand{\Hrule}{\rule{.5\linewidth}{0mm}~\rule{.5\linewidth}{0.5mm}}
\titleformat{\section}[block]{\centering}{\thesection.}{1em}{}
\title{A New Class of Solutions of Combined KdV-mKdV Equation}

\author{\textsc{Sumanta Bandyopadhyay}\thanks{\href{mailto:summanta@gmail.com}{summanta@gmail.com}}\\[2mm]
\normalsize Department of Physics, Washington University, St. Louis, MO 63160, U.S.A  \\\vspace{-5mm}}
\date{}

\begin{document}\maketitle

\begin{abstract}
\noindent In this work, the exact solutions for combined KdV-mKdV generalized equation as a linear superposition of Jacobi elliptic functions, $c_n(\xi,m)$, $d_n(\xi,m)$. When $m$ is set to one, the solution matches with well-known hyperbolic solutions of generalized combined KdV-mKdV equation. Similar solution is also derived for the case for time dependent co-efficient. In the latter case the velocity term become time dependent.
\end{abstract}


\begin{multicols}{2}
\section{Introduction}
The study of nonlinear evolution equations, especially the search for its explicit exact solution, plays an important role in soliton theory. In recent years, various effective methods have been introduced, such as Jacobi elliptic function expansion method [\cite{Hong,HuangLiu}], Darboux transformation [\cite{HuTangLouLiu}], Backlund transformation [\cite{LuHong}], extended sine-cosine method [\cite{TascanBekir}], extended tahh-function mentod [\cite{Fan}] etc. This methods have been extensively used in solving different soliton equations. 

KdV-mKdV equation is one of the most popular equation in soliton physics and appear in many practical scenarios like thermal pulse, wave propagation of bound particle[\cite{Wadati,MNB}] etc. The generalized combined KdV-mKdV equation is given by,
\begin{equation}
u_t+auu_x+bu^2u_x+du_{xxx}=0
\end{equation}\label{e1}
Where $a,b,d$ are constants. Due to the immense uses in practical purposes and popularity in soliton physics this generalized equation is being thoroughly studied and several exact solutions has been proposed for this equation [\cite{Zhang} - \cite{Zhenya}]. This paper discusses a new class of solution of generalized combined KdV-mKdV equation in terms of linear combination of Jacobi elliptic functions. For the same of completeness, the next section gives a brief description of Jacobi elliptic method. In the next section, this method is applied to find out a special class of solution of the eq. \ref{e1}. Fourth section discusses a case, where coefficient of $u_t$ is time dependent.
\section{A Generalized Approach based on Jacobi Elliptic Function}
Consider a  nonlinear Ordinary differential equation,
\begin{equation}
F(u,u_x,u_{xx},\dots)=0
\end{equation}\label{e0}
Now let us assume a polynomial solution of this problem in terms of new variable $y(x)$ in the following form,
$$u(x)=\sum_{i=1}^{n}y^{i-1}(A_i\sqrt{1-y^2}+B_i\sqrt{1-my^2})+A_0$$
Where $y(x)$ follows certain differential equation, in our case it follows Jacobi elliptic algebra where y(x) is idenfied as sn(x,m) and follows the differential equation as written below,
\begin{equation}
y'(x)=\sqrt{1-y^2}\sqrt{1-my^2}
\end{equation}
So our proposed solution can be rewritten as,
\begin{eqnarray}\nonumber
u(x)=\sum_{i=1}^{n}sn(x,m)^{i-1}(A_icn(x,m)\\+B_idn(x,m))+A_0
\end{eqnarray}\label{e2}
Now, inserting this solution to eq. \ref{e0} gives a power series in terms of sn(x,m),cn(x,m),dn(x,m). From linear independence we will set coefficient of each term to zero. This will result in a set of overdetermined nonlinear equations. If after solving them we get a nontrivial values for $(A_i,B_i)$ then we will demand an exact solution to the ODE equation.
\section{New Solitary Wave Solution of KdV-mKdV Equation}
The previous section discusses an approach to solve a nonlinear ODE. Eq. \ref{e1}, however, is a partial differential equation. Now, in eq. \ref{e1} if one set $b=0$ (or $a=0$) the equation will be transformed to KdV(or mKdV) equation. In either cases the solution are well studied and they are function of $(x-vt)$. Where $v$ is the velocity. Here in general case  let us assume assume a soultion as a function of $(x-vt)$. Now let us redefine $x=(x-vt)$, and rewrite the eq. \ref{e1} as follows,
\begin{equation}
-vu'+auu'+bu^2u'+du'''=0
\end{equation}
Where $u'$ denotes total derivative w.r.t. new $x$. This equation looks like, eq.\ref{e0}. We will take the first order term as the ansatz in this equation, i.e,
\begin{equation}
u(x)=A cn(x,m)+B dn(x,m)+ D
\end{equation}
Where the constant term $A_0$ is relabeled as $D$. Using this ansatz in the eq. \ref{e1} we get seven algebraic equations to solve.
\end{multicols}
\noindent\HRule
\begin{eqnarray}
a+2bD +am +2bDm = 0\label{e3}\\a+2bD = 0\\2 A^2 b B + A^2 b B m +b B^3 m - 4 B d m + a B D m + b B D^2 m - B d m^2- B m v = 0\\3 A^2 b + b B^2 m- 6 d m= 0\\A^2b+bB^2-d+aD+bD^2+ 2 b B^2 m - 4dm -v = 0\\ A^2 b + 3b B^2 m - 6d m = 0\\a A^2+ 2 A^2 b D + a B^2 m + 2 b B^2 D m = 0\label{e4}
\end{eqnarray}
\noindent\Hrule
\begin{multicols}{2}
Solving eq. \ref{e3}-\ref{e4} we get, 
\begin{eqnarray}
v=\frac{2bd(1+m)-a^2}{4b},~~~D=-\frac{a}{2b}\\A=\pm\sqrt{\frac{3dm}{2b}},~~~B=\pm\sqrt{\frac{3d}{2b}}
\end{eqnarray}
When $a=0$ and $m=1$ we will retrieve the well known sech solution of mKdV equation.In order to have a real solution $m$ must be positive and $b$, $d$ must have same sign. Sign of $a$ does not effect velocity or the coefficient of $cn(x,m)$ and $dn(x,m)$ but changes the base correction constant, $D$. The explicit solution of the eq. \ref{e1} will be,
$$\pm\sqrt{\frac{3d}{2b}}(\sqrt{m}cn(x-vt,m)\pm dn(x-vt,m))-\frac{a}{2b}$$
i.e, dpending on $\pm$ sign the solutions can be separated into four different classes,
\begin{eqnarray}
AB>0,D<0~~~(a,b~same~sign)\\AB>0,D>0~~~(a,b~different~sign)\\AB<0,D<0~~~(a,b~same~sign)\\AB<0,D>0~~~(a,b~different~sign)
\end{eqnarray}
\section{KdV-mKdV Equation with Time-dependent Coefficient}
Last section shows, there is a solution of combined KdV-mKdV equation if the coefficient of $u^2u_x$ and $u_{xxx}$ has same sign. Now, let us discuss the case where co-efficient of $u_t$ is time dependent. With time dependent co-efficient eq. \ref{e1} will look like,
\begin{equation}
f(t)u_t+auu_x+bu^2u_x+du_{xxx}=0\label{e7}
\end{equation}
Function $f(t)$ can be any arbitrary nonvanishing function of time. Let us multiply eq. \ref{e7} by $h(t)$. Where, $h(t)f(t)=1$. Now the eq. \ref{e7} looks exactly like eq. \ref{e1}, except coefficient of the space derivative terms are multiplied by $h(t)$. Let us look into this equation with the following ansatz,
\begin{equation}\small{u(x,t)=A(t)cn(\xi(x,t))+B(t)dn(\xi(x,t))+D(t)}\end{equation}
Where $\xi(x,t)=x-v(t)t$. With this ansatz, the time derivative part gives some extra contributions.
\begin{eqnarray}\nonumber
u_t=A_tcn(\xi(x,t))+B_tdn(\xi(x,t))+D_t-\\\nonumber (A(t)cn'(\xi(x,t))+B(t)dn'(\xi(x,t)))(v(t)+v_tt)
\end{eqnarray}
Where $cn',dn'$ denote derivatives of $cn,dn$ with respect to $\xi$. Terms consisting of space derivative will be same as the previous case. Hence, we will get the following ordinary differential equation. 
\begin{equation}
-(v+v_tt)u'+h(t)(auu'+bu^2u'+du''')=U(x,t)
\end{equation}
Where, 
\begin{equation}
U(x,t)=A_tcn(\xi(x,t))+B_tdn(\xi(x,t))+D_t
\end{equation}
Now, we get following equations,
\end{multicols}
\noindent\HRule
\begin{eqnarray}
A_t=0,B_t=0,D_t=0\\
a+2bD +am +2bDm = 0\label{e5}\\a+2bD = 0\\h(t)(2 A^2 b B + A^2 b B m +b B^3 m - 4 B d m + a B D m + b B D^2 m - B d m^2)- B m (v+v_tt = 0\\3 A^2 b + b B^2 m- 6 d m= 0\\h(t)(A^2b+bB^2-d+aD+bD^2+ 2 b B^2 m - 4dm) -(v+v_tt) = 0\\ A^2 b + 3b B^2 m - 6d m = 0\\a A^2+ 2 A^2 b D + a B^2 m + 2 b B^2 D m = 0\label{e6}
\end{eqnarray}
\noindent\Hrule
\begin{multicols}{2}
From the first set of equations imply $A,B,D$ are constants. Which is consistent with the rest of the equations. Soving those equations will gives them to be same as the time independent case. 
\begin{equation}
A=\pm\sqrt{\frac{3dm}{2b}},~~B=\pm\sqrt{\frac{3d}{2b}},~~D=-\frac{a}{2b}
\end{equation}
Again, real soultions will exist if $b$ and $d$ has same sign. In this case also one gets different classes of solutions depending on sign of $a$ and $b$. However, the velocity term is no longer constant in this case. It is given by,
\begin{equation}
v=\frac{2bd(1+m)-a^2}{4b}e^{-t}\Bigg(\int^t\frac{dse^{s}}{f(s)}+v_0\Bigg)
\end{equation}
$v_0$ is an arbitrary constant, determined by the boundary condition. 
\section{Conclusion}
Jacobi elliptic function expansion method is implemented to find four new classes of solitary wave solution of general KdV-mKdV equation for both constant and time dependent coefficient cases. The solutions are just linear combination of Jacobi elliptic funtion and has not reported before according to the best of the author's knowledge.
\section{Acknowledgement}
The author would like to thank Prasanta K. Panigrahi for motivating the author to study this problem.
\end{multicols}
\begin{multicols}{2}


\end{multicols}


\begin{thebibliography}{99} 

\bibitem {Hong} B. J. Hong (2009),
\newblock{New Jacobi elliptic funtions solutions for the variable-coeffcient MKdV equation,}
\newblock{\em{Appl. Math. Comput.}}, 215, 2908-2913
\bibitem{HuangLiu} W. H. Huang, Y. L. Liu (2009),
\newblock{Jacobi elliptic function solutions of the Ablowitz-Ladik discrete nonlinear Schrodinger system}, \newblock{\em{Chaos Soliton Fract}}, 40, 786-792
\bibitem{LuHong} D. C. Lu, B. J. Hong (2006), \newblock{Backlund transformation and n-soliton-like solutions to the combined KdV-Burgers equation with variable coeffcient},
\newblock{\em{Int. J. Nonlinear Sci.}}, 1(2), 3-10
\bibitem{Fan} E. G. Fan (2000), \newblock{Extended tanh-function method and its applications to nonlinear equations}, \newblock{\em{Phys. Lett. A.}}, 277, 212-220
\bibitem{HuTangLouLiu} H. C. Hu, X. Y. Tang, S. Y. Lou, Q. P. Liu (2004), \newblock{Variable separation solutions obtained from Darboux transformations for the asymmetric Nizhnik-Novikov-Veselov system}, \newblock{\em{Chaos Soliton Fract}}, 22(2), 327-334
\bibitem{TascanBekir} F. L. Tascan, A. Bekir (2009), \newblock{Analytic solutions of the (2+1)-dimensional nonlinear evolution equations
using the sine-cosine method}. \newblock{\em{Appl. Math. Comput.}}, 215, 3134-3139
\bibitem{Wadati}Wadati M.(1975),\newblock{Wave propatation in nonlinear lattice}, \newblock{\em{J Phys. Soc. Jpn.}}, 38: 673
\bibitem{MNB} Mohamad MNB (1992), \newblock{Exact solutions to the combined KdV and MKdV equation}, \newblock{\em{Math. Meth. Appl. Sci.}}, 15: 73
\bibitem{Zhang} Zhang, J. (1998), \newblock{New solitary wave solution of the combined KdV and mKdV equation}, \newblock{\em{International journal of theoretical physics}}, 37(5), 1541-1546.
\bibitem{LuShi}Lu, D., Shi, Q. (2010),\newblock{New Solitary Wave Solutions for the Combined KdV-MKdV Equation}, \newblock{\em{Journal of Information \& Computational Science}}, 7(8), 1733-1737
\bibitem{Jafari}Mousavian, S. R., Jafari, H., Khalique, C. M., \& Karimi, S. A.(2011), \newblock{New exact-analytical solutions for the mKdV equation}, \newblock{\em{The Journal of Mathematics and Computer Science}}, 2(3), 413-416
\bibitem{Sierra}Sierra, C. A. G., Molati, M., \& Ramollo, M. P. (2012), \newblock{Exact Solutions of a Generalized KdV-mKdV Equation}, \newblock{\em{International Journal of Nonlinear Science}}, 13(1), 94-98
\bibitem{Zhenya} Zhenya, Y., \& Hongqing, Z. (2001),\newblock{Explicit exact solutions for the generalized combined KdV and MKdV equation}, \newblock{\em{Applied Mathematics-A Journal of Chinese Universities}}, 16(2), 156-160.
\end{thebibliography}
\end{document}